\documentclass{aa}
\usepackage{graphicx}
\usepackage{txfonts}
\newcommand{\etal}{et\,al.\ }
\newcommand{\logg}{\mbox{$\log g$}}
\newcommand{\Teff}{\mbox{$T_\mathrm{eff}$}}

\newcommand{\lppr}{\stackrel{<}{\scriptstyle \sim}}
\newcommand{\lappr}{\raisebox{-0.4ex}{$\lppr $}}

\begin{document}
   \title
   {VLT spectroscopy and non-LTE modeling of the C/O-dominated accretion disks in two
   ultracompact X-ray binaries\thanks
{Based on observations made with ESO Telescopes at the Paranal Observatory under programme ID 72.D-0013(A).}
   }
 
   \author{K. Werner$^1$, T. Nagel$^1$, T. Rauch$^{1}$, N.~J. Hammer$^{2,1}$, \and S. Dreizler$^3$}
   \offprints{K. Werner}
   \mail{werner@astro.uni-tuebingen.de}
 
   \institute
    {
     Institut f\"ur Astronomie und Astrophysik, Universit\"at T\"ubingen, Sand 1, 72076 T\"ubingen, Germany
\and
Max-Planck-Institut f\"ur Astrophysik, Karl-Schwarzschild-Stra\ss e 1, 85741 Garching, Germany
\and
Institut f\"ur Astrophysik, Universit\"at G\"ottingen, Friedrich-Hund-Platz 1, 37077 G\"ottingen, Germany
}
    \date{Received xxx / Accepted xxx}
   \authorrunning{K. Werner et al.}
   \titlerunning{VLT spectroscopy of LMXBs 4U\,0614+091 and 4U\,1626-67}
   \abstract{}
{We present new medium-resolution high-S/N optical spectra of the ultracompact
low-mass X-ray binaries 4U\,0614+091 and 4U\,1626-67, taken with the ESO Very
Large Telescope. They are pure emission line spectra and the lines are
identified as due to \ion{C}{ii-iv} and \ion{O}{ii-iii}.}
{Line identification is corroborated by first results from modeling the disk
spectra with detailed non-LTE radiation transfer calculations. Hydrogen and
helium lines are lacking in the observed spectra.}
{Our models confirm the deficiency of H and He in the disks. The lack of neon
lines suggests an Ne abundance of less than about 10 percent (by mass), however,
this result is uncertain due to possible shortcomings in the model atom. These
findings suggest that the donor stars are eroded cores of C/O white dwarfs with
no excessive neon overabundance. This would contradict earlier claims of Ne
enrichment concluded from X-ray observations of circumbinary material, which
was explained by crystallization and fractionation of the white dwarf core.}
{}
   \maketitle

\section{Introduction}

Low-mass X-ray binaries (LMXBs) consist of a neutron star or black-hole accretor
and a low-mass donor star (M\,$\lappr$\,1\,M$_\odot$). Of particular interest
are those systems with orbital periods P$_{\rm orb}$\,$\lappr$\,80\,min, which is the
minimum period for LMXBs with hydrogen-rich main sequence donors. In these
ultracompact binaries (UCBs) the mass donor must be a non-degenerate
hydrogen-deficient star or a white dwarf (e.g.\ Verbunt \& van den Heuvel
1995). Currently eight such systems with measured orbital periods (P$_{\rm
orb}$=11-50\,min) are known (Ritter \& Kolb 2003).

\begin{figure*}[tbp]
\includegraphics[width=\textwidth]{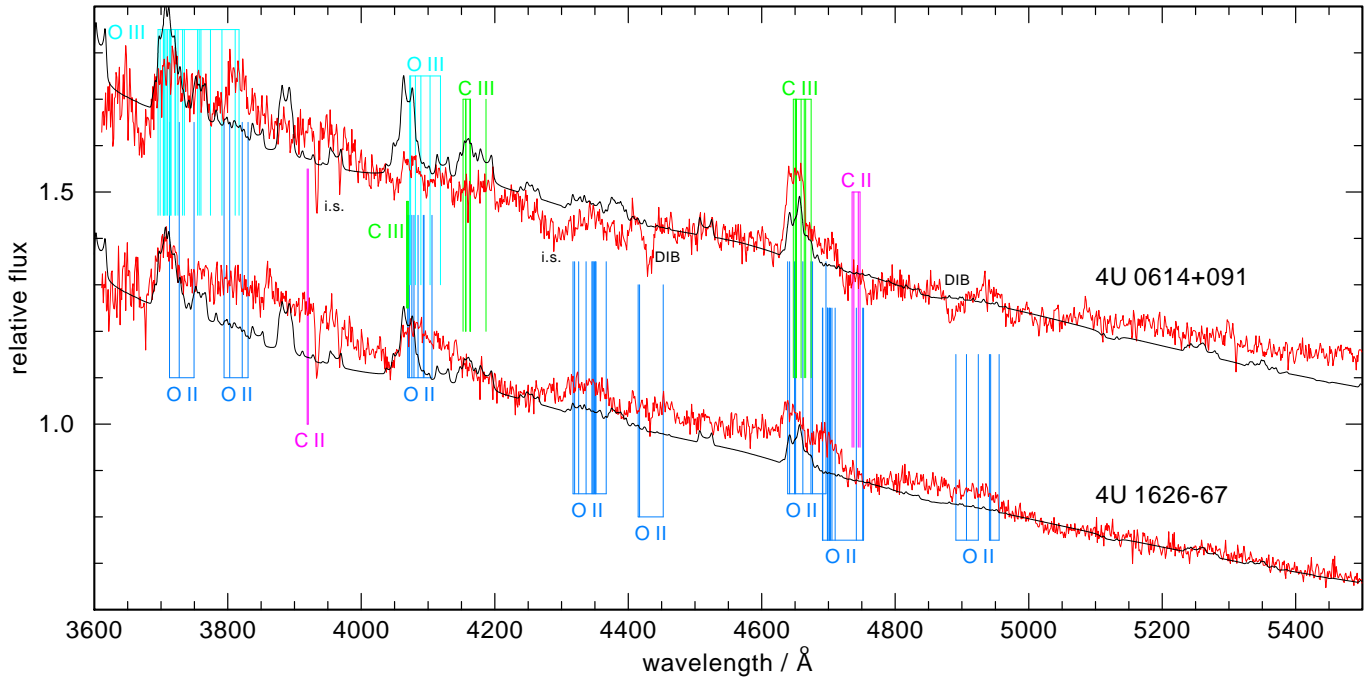}
  \caption[]{Blue spectra of 4U\,0614+091 and 4U\,1626-67. H and He lines are
  lacking. The emission lines are identified as \ion{O}{ii-iii} and
    \ion{C}{ii-iii} (see also Table\,\ref{lines_tab}). All absorption features are
    of interstellar (i.s.) origin: \ion{Ca}{ii} 3934/3968~\AA, the 4300~\AA\ CH band,
    and two diffuse interstellar bands (DIB). Plotted over the two observed spectra 
is a synthetic accretion disk spectrum reddened with
    E(B-V)=0.55 and E(B-V)=0.40, respectively.  }
  \label{fig1}
\end{figure*}

\begin{figure*}[tbp]
\includegraphics[width=\textwidth]{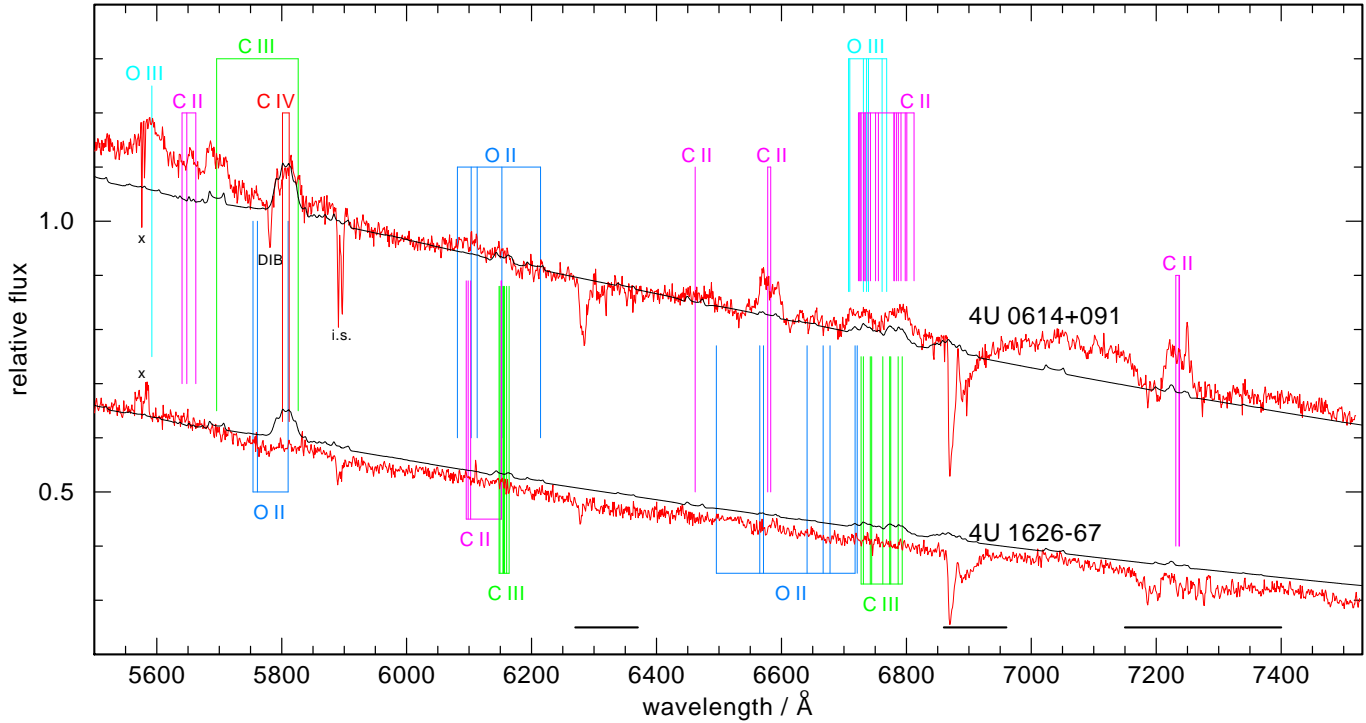}
   \caption[]{Red spectra; 4U\,0614+091 shows emission lines from \ion{C}{ii-iv}
    and \ion{O}{ii-iii}, while the 4U\,1626-67 spectrum is virtually featureless. All
    absorption features are either telluric (marked by horizontal bars) or of
    interstellar origin (\ion{Na}{i} 5890/5896~\AA\ and a DIB
    at 5780~\AA). The features at 5575~\AA\ are artifacts (``x'') due to an 
    \ion{O}{i} sky emission line. Overplotted is the synthetic accretion disk
    spectrum like in Fig.\,\ref{fig1}.
}
  \label{fig2}
\end{figure*}

\begin{figure*}[tbp]
\includegraphics[width=\textwidth]{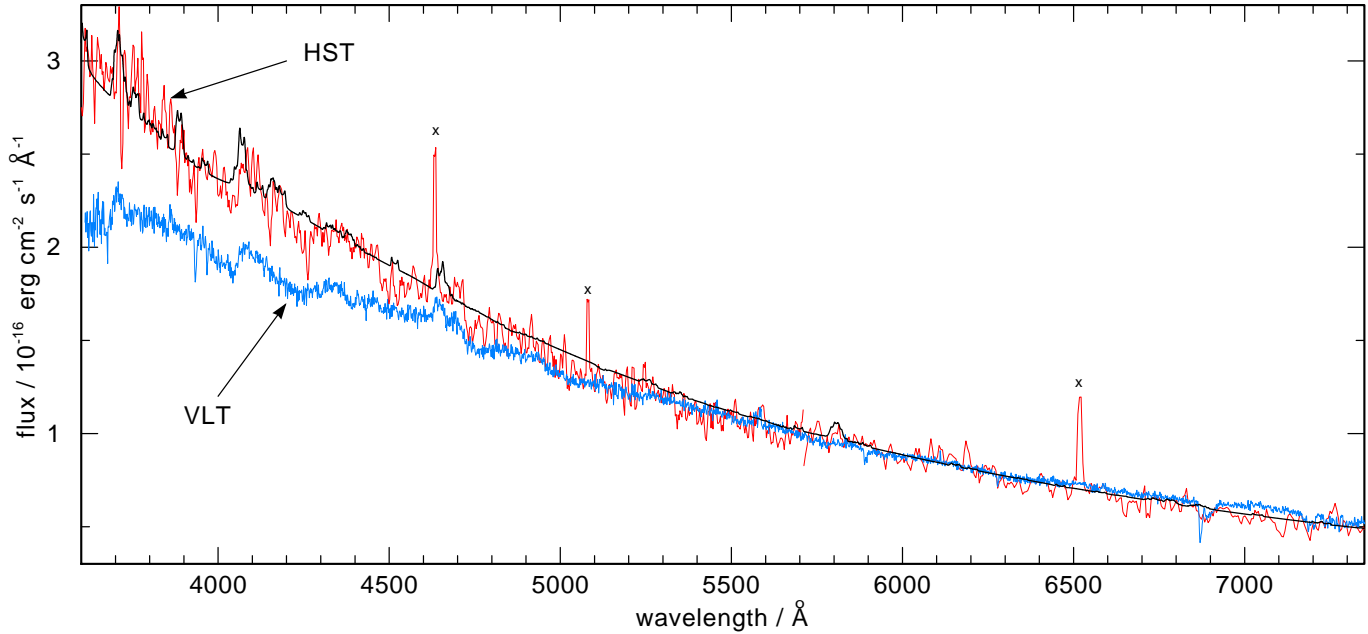}
  \caption[]{Complete VLT spectrum of 4U\,1626-67 compared to a spectrum taken with
    HST. The VLT spectrum, scaled to the HST spectrum at 5600~\AA, is flatter
    possibly due to problems in absolute flux calibration. The HST spectrum has
     a poorer resolution and S/N-ratio, although some emission features at
    $\lambda<5000$~\AA\ can be recognized in both datasets. Artificial emission
    spikes are marked by ``x''. Overplotted is the synthetic
    accretion disk spectrum, like in previous figures, but now reddened with
    E(B-V)=0.20 to fit the continuum shape of the HST spectrum. }
  \label{fig3}
\end{figure*}

\begin{figure*}[tbp]
\includegraphics[width=\textwidth]{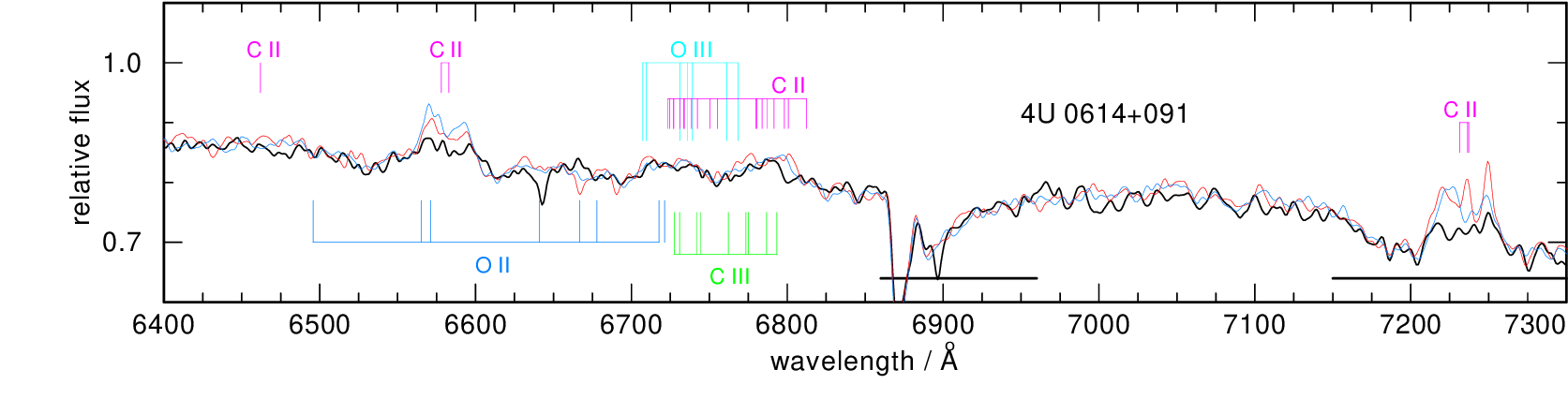}
  \caption[]{Detail from three single spectra of 4U\,0614+091. Thick line:
    Spectrum taken on Nov. 03, 2003. Two thin lines: Spectra taken consecutively
    on Dec. 15, 2003 (see Table~\ref{obs_tab}). The strength of the two strongest emission features, the
    \ion{C}{ii} multiplets at 6580~\AA\ and 7235~\AA, show an increase
    over a time interval of several weeks.
    }
  \label{fig4}
\end{figure*}

In the recent past, the existence of a group of five ultracompact systems with
neon-rich white dwarf donors has been claimed based on X-ray spectral properties
(Schulz \etal 2001, Juett \etal 2001, Juett \& Chakrabarty 2003). The donors are
then C/O or even O/Ne/Mg white dwarfs that have transferred a significant
fraction of their mass to the neutron star, in contrast to the usual wisdom that
the donors are the remains of He white dwarfs. This has caused new explorations
of the formation of these systems (e.g.\ Yungelson \etal 2002). Our
motivation for studying these systems is that the stripped donor stars offer the
possibility to probe the interior composition of white dwarfs, which depends on
the interplay of gravitational settling and crystallization of chemical elements.

Three of these five Ne-rich systems belong to the above-mentioned eight UCBs
with measured orbital periods, while two of them are believed to be UCBs because
of their similar optical and X-ray properties. The class of Ne-rich UCBs
consists of:
\begin{tabbing}
(1) 4U\,1626-67  \quad  \= (P$_{\rm orb}$=41\,min)\\
(2) 4U\,0614+091        \>                        \\
(3) 2S\,0918-549        \>                        \\
(4) 4U\,1543-624        \> (P$_{\rm orb}$=18\,min, Wang \& Chakrabarty 2004)\\
(5) 4U\,1850-087        \> (P$_{\rm orb}$=20\,min)
\end{tabbing}
And 4U\,1626-67 may be regarded as the prototype of this class. The donor's Ne-rich
C/O-WD nature is derived from X-ray and ultraviolet spectra that exhibit
double-peaked emission lines that obviously stem from the accretion disk (Schulz
\etal 2001, Homer \etal 2002).

The close relation of the other four objects (2)\,--\,(5) to 4U\,1626-67 was
based on extraordinary high Ne/O abundance ratios (when compared to the ISM
value) measured from ASCA spectra of (3)--(5) and a Chandra  spectrum of
(2). The spectra exhibit photoelectric absorption edges of neutral O and Ne in
the interstellar medium (ISM) along the line-of-sight, which is suspected to
originate from expelled material close to the binary systems. New X-ray
spectroscopic observations of (3) and (4) obtained with Chandra and XMM-Newton
confirmed the ASCA results, although different values for the Ne/O enrichment
have been obtained from a Chandra and a XMM-Newton spectrum of (4) (Juett \&
Chakrabarty 2003). In contrast to the ASCA measurement of (5), recent XMM-Newton
and Chandra observations found no evidence of an unusual Ne/O ratio (Sidoli
\etal 2005, Juett \& Chakrabarty 2005). These apparently contradictory results
can be attributed to a variable \ion{Ne}{i}/\ion{O}{i} ratio due to changes in
the ionisation structure in the measured absorption columns that, however, are
not understood (Juett \& Chakrabarty 2005). Hence, this means that the measured
Ne/O ratio does not reflect the donor composition. Although the orbital periods
of these systems are below 80\,min (or at least believed to be so small), this
does not necessarily mean that they contain C/O donors. For example, the X-ray
burst properties of the 4U\,1820-30 (P$_{\rm orb}$=11\,min) suggest an He-WD
donor in that ultracompact system (e.g.\ Strohmayer \& Brown 2002). Another
example are the relatively short X-ray bursts observed from (3) that even
suggest H-burning of material accreted onto the neutron star (NS) (Jonker \etal
2001), while optical spectra suggest a C/O WD donor (Nelemans \etal 2004 and this
work). A possible explanation is that the heavy elements (C, O, Ne) undergo
spallation during accretion leaving H and He nuclei for thermonuclear burning on
the NS (Bildsten \etal 1992). In a recent paper In't Zand \etal (2005)
conclude that the Ne/O ratio and the X-ray burst properties are all best
explained with an He-rich donor.

In conclusion, the only way to confirm that the
four systems (2)--(5) -- besides the prototype (1) -- indeed contain C/O-rich donors, perhaps enriched with Ne,
is by UV and/or optical spectroscopy. Nelemans \etal (2004) show that the
optical spectra of (2)--(4) are devoid of hydrogen and helium emission lines,
which are usually seen in H- or He-rich accreting systems. Their spectra exhibit
low-ionisation C and O emission lines, most prominent in the brightest of these
objects, 4U\,0614+091.  In this paper we present new optical spectra of this
system that cover a larger wavelength interval. They confirm the earlier
conclusion by Nelemans \etal (2004) that the emission lines arise from the
C/O-dominated material that is probably located in the accretion disk and not in
the X-ray heated face of the white-dwarf donor. We also present a first
detailed optical spectrum of the prototype 4U\,1626-67, which mainly shows weak
oxygen emission lines and proves the H- and He- deficiency in this system,
too. Nelemans \& Jonker (2005) also performed VLT observations of this system
(also in spring 2004) with a similar setup. They show a section of their
spectrum and emphasize the similarity with 4U\,0614+91.

In addition we present results from first attempts to model the observed spectra
with non-LTE accretion-disk models. We derive upper limits for the H and He
abundances and investigate the formation of neon lines. Detailed C and O
line-formation calculations can already qualitatively explain the observed
emission lines. The ultimate goal is to derive detailed abundances and other
disk parameters from the observed line profiles.

The paper is organised as follows. We describe our observations in the following
section. We then present our line identifications in Sect.~3. Section~4
contains a description of our disk model calculations and we summarise and
conclude in Sect.~5.

\section{Observations}

We obtained optical medium-resolution long-slit spectra of 4U\,1626-67 and
4U\,0614+091. The V magnitude of both systems is 18.5 (Ritter \& Kolb 2003). We
used the FORS1 spectrograph attached to UT1 of ESO's Very Large Telescope
(VLT) on Paranal in Chile. The slit width was 1\arcsec. We used two grisms (600B
and 600R, the latter in combination with order separation filter GG435) and
obtained spectra covering the regions 3600--6000~\AA\ and 5400--7500~\AA\ with a
mean dispersion of 1.20 and 1.07~\AA~pix$^{-1}$, respectively. Observations were
performed in service mode between Nov.~2003 and Mar.~2004. Each target was
observed at least twice, see Table~\ref{obs_tab} for details. For 4U\,1626-67
the exposure time covers almost one orbital period so that any respective spectral
variation is smeared out. The data were
processed through the standard ESO reduction pipeline. In addition we performed
a flux calibration using observations of the DA white dwarf EG\,274 with the same
instrumental setup. Since this flux standard was observed only once per
wavelength range and since the science targets were mostly observed at different
dates, this provides only a rough absolute flux calibration. For each object the
spectra were co-added to obtain one final spectrum. The blue and red spectra are
displayed in Figs.\,\ref{fig1} and~\ref{fig2}, respectively.

Archival spectra of 4U\,1626-67 taken with the Hubble Space Telescope (HST) and
the STIS instrument cover the complete UV/optical wavelength range
(1150--10\,000~\AA). The FUV spectrum is described in detail by Homer \etal
(2002) and a first comparison with synthetic accretion disk spectra was
presented by Werner \etal (2004). In Fig.\,\ref{fig3} we compare the optical HST
spectrum taken from the MAST archive with our VLT spectrum on an absolute flux
scale. The blue and red sections of the VLT spectrum were scaled by a factor of
1.65 and 1.45, respectively, to normalise them to the HST flux at 5600~\AA. It
is obvious that the VLT spectrum is not only weaker but also flatter than
the HST spectrum. This can be attributed either to flux calibration problems or
to source variability.

\begin{table}
\caption{Log of the observations.
\label{obs_tab}
}
\begin{tabular}{llccc}
      \hline
      \hline
      \noalign{\smallskip}
Date, UT               & Grism & T$_{\exp}$ & Airmass & Seeing\\
                       &       & (sec)     &         & (arcsec)\\
      \noalign{\smallskip}
     \hline
      \noalign{\smallskip}
 & \multicolumn{2}{c}{4U\,0614+091}\\
2003 Nov. 04, 07:22:08 & 600B & 1735 & 1.22 & 0.73 \\
2003 Nov. 04, 07:54:57 & 600B & 1735 & 1.20 & 0.78 \\
2003 Nov. 03, 07:40:17 & 600R & 1735 & 1.21 & 1.00 \\
2003 Dec. 15, 06:25:42 & 600R & 1735 & 1.28 & 0.52 \\
2003 Dec. 15, 07:00:31 & 600R & 1735 & 1.37 & 0.57 \\
      \noalign{\smallskip}
 & \multicolumn{2}{c}{4U\,1626-67}\\
2004 Mar. 23, 08:22:00 & 600B & 1735 & 1.37 & 0.76 \\
2004 Mar. 23, 09:01:42 & 600B & 1735 & 1.36 & 0.77 \\
2004 Mar. 21, 08:48:53 & 600R & 1735 & 1.36 & 1.58 \\
2004 Mar. 22, 08:19:40 & 600R & 1735 & 1.37 & 0.87 \\
2004 Mar. 22, 09:00:00 & 600R & 1735 & 1.36 & 0.71 \\
2004 Mar. 23, 07:41:51 & 600R & 1735 & 1.40 & 0.57 \\
      \noalign{\smallskip}
 & \multicolumn{2}{c}{EG\,274}\\
2004 Mar. 23, 09:48:17 & 600B & 50   & 1.05  & 0.67 \\
2004 Mar. 23, 09:53:25 & 600R & 40   & 1.05  & 0.67 \\
           \noalign{\smallskip}
\hline
     \end{tabular}
\end{table}

\begin{table*}
\caption{Emission features observed in the spectra of 4U\,1626-67 and
  4U\,0614+091 and suggested line identifications. A plus sign in parentheses
  denotes an uncertain detection.
\label{lines_tab}
}
\begin{tabular}{ccclcc}
      \hline
      \hline
      \noalign{\smallskip}
Feature&\multicolumn{2}{c}{Observed in}&Ion&Transition    & Wavelength \\
(\AA)  &  4U\,0614+091 & 4U\,1626-67 &   &              &   (\AA)    \\
      \noalign{\smallskip}
     \hline
      \noalign{\smallskip}
3720&+&+& \ion{O}{iii} & 3p $^3$P -- 3d $^3$D$^{\rm o}$ & 3704--3732 \\
    & & & \ion{O}{iii} & 3s $^5$P -- 3p $^5$D$^{\rm o}$ & 3695--3735 \\
    & & & \ion{O}{ii}  & 3s $^4$P -- 3p $^4$S$^{\rm o}$ & 3713--3749 \\ 
      \noalign{\smallskip}
3815&+& & \ion{O}{ii}  & 3p $^2$P$^{\rm o}$ -- 4s $^2$P & 3794--3830 \\
    & & & \ion{O}{iii} & 3p $^1$D -- 3d $^1$P$^{\rm o}$ & 3817       \\
    & & & \ion{O}{iii} & 3s $^3$P$^{\rm o}$ -- 3p $^3$D & 3755--3811 \\
      \noalign{\smallskip}
3920& &(+) & \ion{C}{ii}  & 3p $^2$P$^{\rm o}$ -- 4s $^2$S & 3919, 3921 \\
      \noalign{\smallskip}
4075&+&+& \ion{O}{ii}  & 3p $^4$D$^{\rm o}$ -- 3d $^4$F & 4070--4111 \\
    & & & \ion{C}{iii} & 4f $^3$F$^{\rm o}$ -- 5g $^3$G & 4068--4070 \\
    & & & \ion{O}{iii} & 3s $^3$P -- 3p $^3$D$^{\rm o}$ & 4073--4119 \\
      \noalign{\smallskip}
4180&+& & \ion{C}{iii} & 4f $^1$F$^{\rm o}$ -- 5g $^1$G & 4187 \\
    & & & \ion{C}{iii} & 2p3p $^3$D -- 2s5f $^3$F$^{\rm o}$ & 4153--4163 \\
      \noalign{\smallskip}
4345&+&+& \ion{O}{ii}  & 3s $^4$P -- 3p $^4$P$^{\rm o}$ & 4317--4367 \\
    & & & \ion{O}{ii}  & 3s' $^2$D -- 3p' $^2$D$^{\rm o}$ & 4347--4351 \\
      \noalign{\smallskip}
4410&+&(+)&\ion{O}{ii}  & 3s $^2$P -- 3p $^2$D$^{\rm o}$ & 4415--4452 \\
      \noalign{\smallskip}
4650&+&+& \ion{C}{iii} & 3s $^3$S -- 3p $^3$P$^{\rm o}$ & 4647--4651 \\
    & & & \ion{C}{iii} & 3s $^3$P$^{\rm o}$ -- 3p $^3$P & 4651--4674 \\
    & & & \ion{O}{ii}  & 3s $^4$P -- 3p $^4$D$^{\rm o}$ & 4639--4696 \\ 
      \noalign{\smallskip}
4710&+&+& \ion{O}{ii}  & 3p' $^2$P$^{\rm o}$ -- 3d' $^2$P & 4691--4702 \\
    & & & \ion{O}{ii}  & 3p $^2$D$^{\rm o}$ -- 3d $^2$F & 4699--4742 \\
    & & & \ion{O}{ii}  & 3p' $^2$D$^{\rm o}$ -- 3d' $^2$F & 4698--4703 \\
    & & & \ion{O}{ii}  & 3p $^2$D$^{\rm o}$ -- 3d $^4$D & 4710--4753 \\
      \noalign{\smallskip}
4745&(+)& & \ion{C}{ii}  & 2s2p$^2$ $^2$P -- 2s$^2$3p $^2$P$^{\rm o}$ & 4735--4747 \\
      \noalign{\smallskip}
4940&+&(+)& \ion{O}{ii}  & 3p $^2$P$^{\rm o}$ -- 3d $^2$D & 4941--4956 \\
    & & & \ion{O}{ii}  & 3p $^4$S$^{\rm o}$ -- 3d $^4$P & 4891--4925 \\
      \noalign{\smallskip}
5590&+&(+)& \ion{O}{iii} & 3s $^1$P$^{\rm o}$ -- 3p $^1$P & 5592       \\
      \noalign{\smallskip}
5650&+& & \ion{C}{ii}  & 3s $^4$P$^{\rm o}$ -- 3p $^4$S & 5641--5662 \\
      \noalign{\smallskip}
5695&+& & \ion{C}{iii} & 3p $^1$P$^{\rm o}$ -- 3d $^1$D & 5696       \\
      \noalign{\smallskip}
5810&+& & \ion{O}{ii}  & 2p$^4$ $^2$P -- 3p' $^2$D$^{\rm o}$ & 5754--5810 \\
    & & & \ion{C}{iv}  & 3s $^2$S -- 3p $^2$P$^{\rm o}$ & 5801, 5812 \\
    & & & \ion{C}{iii} & 2s4d $^1$D -- 2p3d $^1$F$^{\rm o}$ & 5826 \\
      \noalign{\smallskip}
6100&+& & \ion{C}{ii}  & 3p $^2$P -- 3d $^2$D$^{\rm o}$ & 6095--6103 \\
    & & & \ion{O}{ii}  & 3p $^2$P$^{\rm o}$ -- 3s $^2$S & 6081, 6103 \\
    & & & \ion{O}{ii}  & 4s $^4$P -- 3s $^4$S$^{\rm o}$ & 6113, 6153 \\
%    & & & \ion{O}{iii} & 3s $^3$P$^{\rm o}$ -- 2p$^4$ $^3$P & 5980--6200 \\  %   too weak
      \noalign{\smallskip}
6150&+& & \ion{C}{iii} & 2s4d $^3$D -- 2p3d $^3$D$^{\rm o}$ & 6148--6164 \\
    & & & \ion{C}{ii}  & 4d $^2$D -- 6f $^2$F$^{\rm o}$ & 6151       \\
    & & & \ion{O}{ii}  & 4s $^4$P -- 3s $^4$S$^{\rm o}$ & 6113, 6153, 6214 \\
      \noalign{\smallskip}
6460&+& & \ion{C}{ii}  & 4f $^2$F$^{\rm o}$ -- 6g $^2$G & 6462 \\
      \noalign{\smallskip}
6580&+& & \ion{C}{ii}  & 3s $^2$S -- 3p $^2$P$^{\rm o}$ & 6578, 6583 \\
    & & & \ion{O}{ii}  & 3d $^2$F -- 4p $^2$D$^{\rm o}$ & 6496, 6565, 6571 \\
      \noalign{\smallskip}
6730&+& & \ion{C}{ii}  & 4d $^2$D -- 6p $^2$P$^{\rm o}$ & 6723       \\
    & & & \ion{C}{ii}  & 3p $^4$D -- 3d $^4$D$^{\rm o}$ & 6725--6755 \\
    & & & \ion{C}{iii} & 3s $^3$P$^{\rm o}$ -- 2p $^3$D & 6727--6773 \\
    & & & \ion{O}{ii}  & 3s $^2$P -- 3p $^2$S$^{\rm o}$ & 6641, 6721 \\
    & & & \ion{O}{ii}  & 3d $^2$P -- 4p $^2$P$^{\rm o}$ & 6667, 6678, 6718 \\
    & & & \ion{O}{iii} & 4d $^3$D$^{\rm o}$ -- 3s $^3$D & 6707--6768 \\
      \noalign{\smallskip}
6790&+& & \ion{C}{ii}  & 3s $^4$P$^{\rm o}$ -- 3p $^4$D & 6780--6812 \\
    & & & \ion{C}{iii} & 2p3d $^3$P$^{\rm o}$ -- 2s6s $^3$S & 6775--6793 \\
      \noalign{\smallskip}
7235&+& & \ion{C}{ii}  & 3p $^2$P$^{\rm o}$ -- 3d $^2$D & 7231--7237 \\
           \noalign{\smallskip}
\hline
     \end{tabular}
\end{table*}

\section{Line identification}

Emission lines from hydrogen and helium are completely lacking in the VLT
spectra of both systems. We detect neither Balmer lines nor lines from
\ion{He}{i} (e.g. 4471 or 5876~\AA) or \ion{He}{ii} (e.g. 4686~\AA), which are
usually seen in optical spectra of interacting binaries. In the following two
sections we discuss the line spectra of both systems in detail. Line
identification is performed ``by eye'' without an equivalent-width estimation.

\subsection{4U\,0614+091}

The spectrum of 4U\,0614+091 shows numerous emission features that can be
assigned to ionized carbon and oxygen, namely \ion{C}{ii-iv} and
\ion{O}{ii-iii}. The detected emission features are listed in
Table\,\ref{lines_tab} together with possible line identifications. Most
features are blends of lines from at least two ions, making interpretation
quite difficult, but some of them are probably unblended. These are the features
at 5590~\AA\ (\ion{O}{iii}), 5650~\AA\ (\ion{C}{ii}), 5695~\AA\ (\ion{C}{iii}),
and 7235~\AA\ (\ion{C}{ii}). The emission features at 4710 and 4940~\AA\ are
perhaps exclusively due to \ion{O}{ii}, but they are blends of four and two
multiplets from this ion, respectively.  Nelemans \etal (2004) present
optical spectra of this star taken with VLT+FORS2 with slightly different
wavelength coverage (4620--8620~\AA). We essentially confirm the detection of
their emission features and line identifications, however, our spectra extend to
shorter wavelengths, down to 3600~\AA, which allows us to detect some additional
emission lines. On the other hand, we do not see four of the weak emission features
identified in the Nelemans \etal (2004) spectra (at $\lambda\lambda$ 5140, 5190,
5280, and 6070~\AA). This could be the consequence of line variability. In
Fig.\,\ref{fig4} we compare our single spectra of 4U\,0614+091 in detail, and it
is possible that the emission strength of the strongest \ion{C}{ii} lines is
variable on a time scale of weeks. In order to assess the significance of this
variability, we simulated several times with different noise the strongest emission features observed on Nov.~3 and
Dec.~15, applying the S/N ratio of the VLT spectra. We find a 2$\sigma$
probability that the variability is real. The line widths correspond to a projected
rotational velocity of the order 1200~km\,s$^{-1}$.

\subsection{4U\,1626-67}

The spectrum of 4U\,1626-67 clearly shows less features than does that of
4U\,0614+091. Only five emission lines can be detected in the blue spectrum (see
Table\,\ref{lines_tab}) and they are seen in both binaries. The red spectrum of
4U\,1626-67 is virtually continuous. The strong \ion{C}{ii} emissions seen in
4U\,0614+091 at 6580~\AA\ and 7235~\AA\ are lacking in 4U\,1626-67. That could
point to a higher ionisation of carbon in 4U\,1626-67, although the
\ion{C}{iv}~5810~\AA\ line is lacking, too. The presence of \ion{O}{ii} lines in
the blue spectrum of 4U\,1626-67 (e.g.\ at 4345~\AA) also contradicts a higher
ionisation, rather, it appears that the differences in the spectra of both
binaries could be assigned to a different C/O ratio, being higher in
4U\,0614+091. The entire set of emission lines in 4U\,1626-67 could be due to
oxygen alone, but in the UV spectrum C is clearly present (Homer \etal
2002). In addition, the line widths appear broader than in 4U\,0614+091,
however, this is difficult to quantify because of possible line blends and an
uncertain identification of the continuum.

All absorption features seen in our optical spectra are either of interstellar
origin or telluric. They are marked in Figs.~\ref{fig1} and~\ref{fig2}.

\section{Exploratory disk models}

We began the construction of accretion disk models to calculate synthetic
spectra and report here on the current state of our work. We use our accretion
disk code AcDc, which is described in detail by Nagel \etal (2004). In essence,
it assumes a radial $\alpha$-disk structure (Shakura \& Sunyaev 1973). Then the
disk is divided into concentric annuli. For each annulus we solve the radiation
transfer equation (assuming plane-parallel geometry) together with the non-LTE
rate equations for the atomic level populations, plus energy- and hydrostatic
equations, in order to calculate a detailed vertical structure. The integrated
disk spectrum is then obtained by co-adding the specific intensities from the
individual annuli, accounting for inclination and Keplerian rotation.

It is not the aim of this paper to present a detailed fit to the emission line
spectra of the two binaries. This requires extensive parameter studies that are
extremely time consuming. We rather choose to calculate synthetic spectra from
selected disk regions in which we believe the physical properties are
representative for the formation of the observed optical spectra. This should at
least give a rough idea of the relative strength of lines from different
ionisation stages and, thus, is primarily thought to put our line
identifications on firm ground. At present we neglect irradiation of the disk
by the neutron star, because it would introduce new free parameters. We expect
that the radial ionisation structure of the disk will be shifted to larger radii
when irradiation is taken into account, but we hope that the relative line
strengths are not affected to the extent that our identifications become
completely wrong. Although the studied systems are strong X-ray sources, we do not see
recombination lines from highly ionised species in the optical spectra as might
be expected. It is conceivable that the outer parts of the disks in which the
optical spectrum arises is shielded from X-ray irradiation by an inflated inner-disk region.

\begin{table}
\caption{Summary of model atoms used in the disk model calculations. For
  each ion we list the number of NLTE levels and the number of line
  transitions. In brackets we give the number of lines in the wavelength
  range covered by our optical spectra after fine-structure splitting for
  detailed line-profile calculations.
\label{levels_tab}}
\small
\begin{tabular}{c c c c}
      \hline
      \hline
      \noalign{\smallskip}
Element & Ion & NLTE levels & Lines\\
  \noalign{\smallskip}
 \hline
      \noalign{\smallskip}
      H  & \scriptsize I          & 10 & 45 (8) \\
         & \mbox{\scriptsize II}  & 1  & --  \\ 
      \noalign{\smallskip}
      He & \scriptsize I          & 29 & 61 (16) \\
         & \mbox{\scriptsize II}  & 14 & 91 (14) \\ 
         & \mbox{\scriptsize III} & 1  & --  \\ 
      \noalign{\smallskip}
      C  & \mbox{\scriptsize I  } & 7  & 4   (0)\\
         & \mbox{\scriptsize II}  & 38 & 160 (42)\\
         & \mbox{\scriptsize III} & 58 & 329 (143)\\
         & \mbox{\scriptsize IV}  & 9  & 17  (2)\\
         & \mbox{\scriptsize V}   & 1  & --  \\
      \noalign{\smallskip}
      O  & \mbox{\scriptsize I  } & 1  & -- \\
         & \mbox{\scriptsize II}  & 29 & 82 (56) \\
         & \mbox{\scriptsize III} & 36 & 42 (47) \\
         & \mbox{\scriptsize IV}  & 11 & 5  (0) \\
         & \mbox{\scriptsize V}   & 6  & 4  (0)\\
         & \mbox{\scriptsize VI}  & 1  & --  \\
      \noalign{\smallskip}
      Ne & \mbox{\scriptsize I  } & 3  & 0  \\
         & \mbox{\scriptsize II}  & 68 & 232 (49) \\
         & \mbox{\scriptsize III} & 4  & 0  (0) \\
         & \mbox{\scriptsize IV}  & 1  & -- \\
           \noalign{\smallskip}
\hline
\normalsize
     \end{tabular}
\normalsize
\end{table}

\begin{figure}[tbp]
\includegraphics[width=80mm]{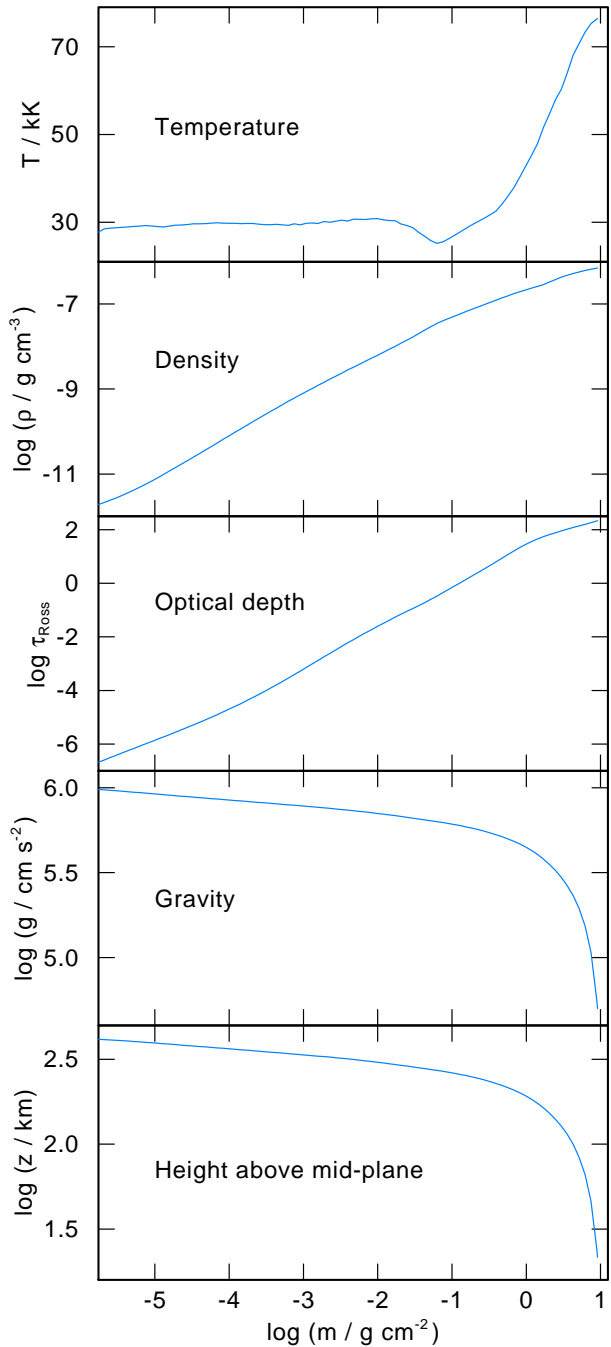}
  \caption[]{Vertical structure of the C/O-disk model at a distance of
    20\,000~km from the neutron star. The emergent disk flux
    at this location corresponds to \Teff=28\,000~K. The physical variables are plotted against the column mass
    measured from the surface towards the mid-plane. 
    }
  \label{fig_structure}
\end{figure}

\begin{figure}[tbp]
\includegraphics[width=79.52mm]{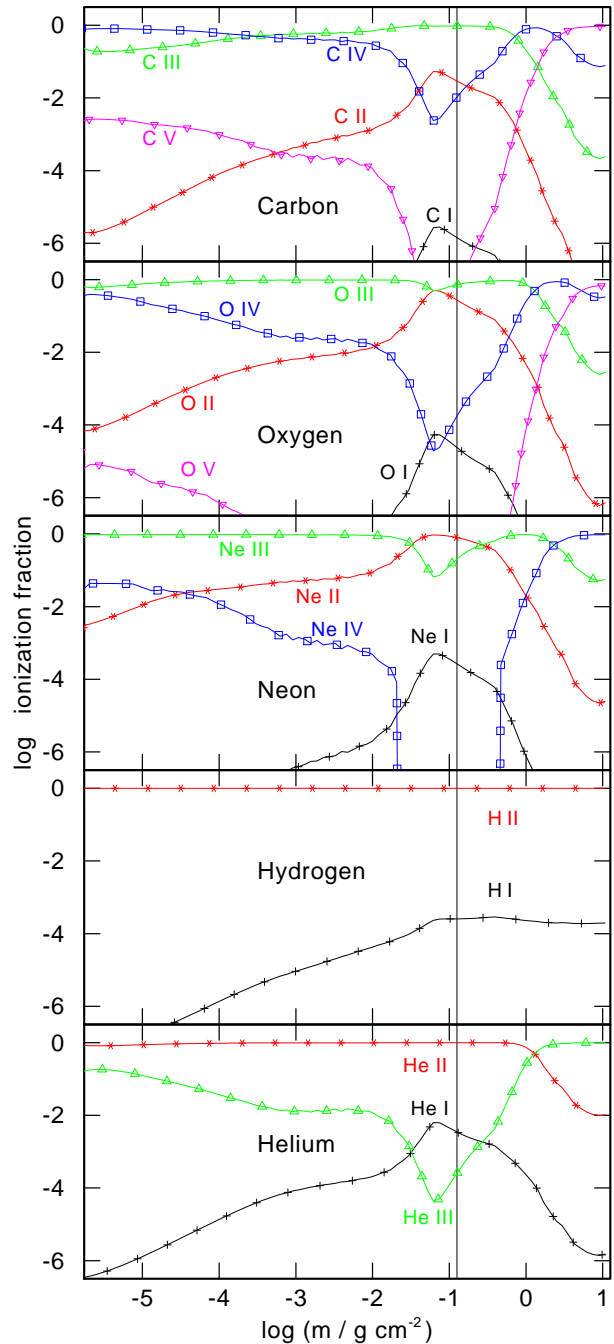}
  \caption[]{Vertical ionisation stratification of chemical elements in the disk models at
    a distance of 20\,000~km from the neutron star. The vertical line drawn at
    $\log m=-0.9$ indicates
    $\tau_{\rm Ross}=1$.
    }
  \label{fig_ion}
\end{figure}

\begin{figure}[tbp]
\includegraphics[width=8cm]{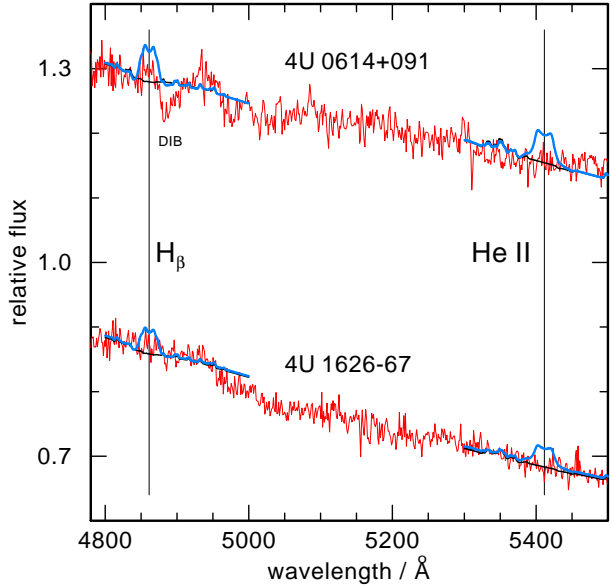}
  \caption[]{Comparison of models with zero and 10\% hydrogen and helium
    content to the observations. From the lack of H$_\beta$ and \ion{He}{ii}
    emission lines in the
    observed spectra, we conclude that the accretion disks are strongly H and He
    deficient. The model spectra were normalised to the local continuum flux.
    }
  \label{fig_hhe}
\end{figure}

\begin{figure}[tbp]
\includegraphics[width=8cm]{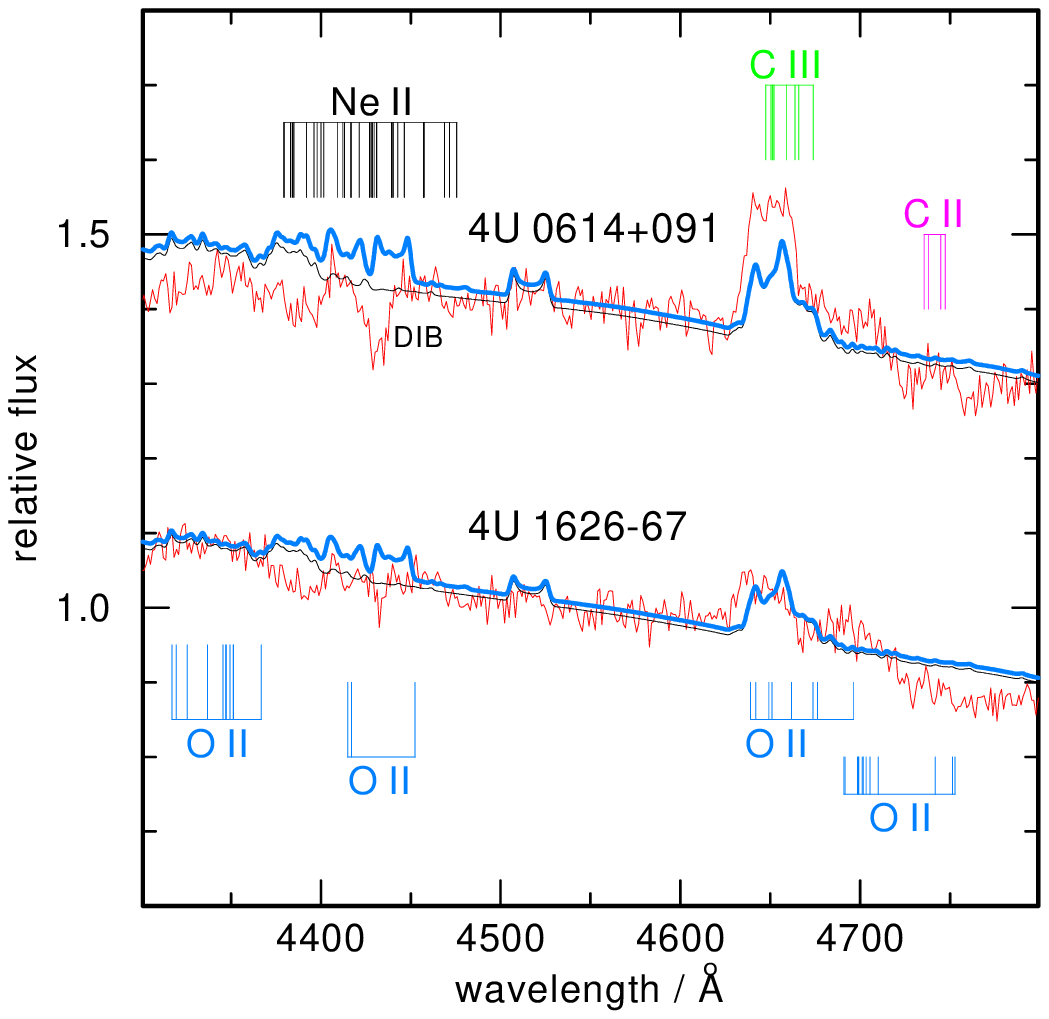}
  \caption[]{Comparison of models with zero and 10\% neon content 
   to the observations. The model including neon exhibits weak \ion{Ne}{ii}
   lines in the region 4350 -- 4450~\AA, which cannot
be detected in the observed spectra.
    }
  \label{fig_neon}
\end{figure}

Since we do not compute the spectrum of the entire disk, we cannot expect to
match the overall observed continuum flux, and we have already pointed out that there
are problems with the absolute flux calibration. We use interstellar reddening
as a free parameter in order to roughly fit the model to the observed flux
level. The applied reddening is given in the figure captions.

The synthetic spectra presented here are based on the following choice of
disk-model parameters. The central object is a neutron star with 10~km radius and a
mass of 1.4~M$_\odot$. The mass accretion rate is $2\cdot
10^{-10}$~M$_\odot$/yr. The disk extends from 1000 to 2000 stellar radii. The
corresponding Keplerian velocities at these radii amount to 4300 and
3000~km\,s$^{-1}$, and the effective temperatures to 47\,000~K and 28\,000~K,
respectively. The disk is divided into five annuli such that \Teff\ decreases
almost linearly with radius steps. The inclination angle is set to 10$\degr$. The
Reynolds number used to parametrise the disk viscosity was set to 10\,000, which
corresponds to $\alpha \approx 0.5$.  The chemical composition is C=50\% and
O=50\% (by mass).

For the opacity and emissivity calculations, it is essential to solve the non-LTE
rate equations with detailed model atoms. Our main emphasis was put on the
\ion{C}{ii-iii} and \ion{O}{ii-iii} ions. The number of non-LTE levels and
radiative line transitions are summarised in Table~\ref{levels_tab}. Level
energies, oscillator strengths, and bound-free cross-sections for
photoionisation are taken from the Opacity Project (Seaton \etal 1994)
TOPbase\footnote{http://vizier.u-strasbg.fr/topbase/topbase.html}. Electron
collisional rates for (de-) excitation, ionisation, and recombination are
computed with the usual approximate formulae. The neon model atom was taken from
Dreizler (1993). For the final spectrum synthesis, fine structure splitting of
atomic levels must be considered; level populations were taken from the models
with appropriate statistical weights. Level energies were obtained from the
NIST\footnote{http://physics.nist.gov/} database. The optical synthetic spectra
finally include a total number of 377 lines from \ion{H}{i}, \ion{He}{i-ii},
\ion{C}{ii-iv}, \ion{O}{ii-iii}, and \ion{Ne}{ii}. They essentially contain most
of the possible C and O line identifications given in Table~\ref{lines_tab},
plus many others that turn out to be too weak to be seen in the rotationally
broadened spectra.

As an example we present the vertical structure of the accretion disk model at a
distance of 2000 stellar radii from the neutron star. The emergent flux at this
location corresponds to \Teff=28\,000\,K. Figure~\ref{fig_structure} shows the
run of several quantities above the disk midplane on a column mass scale
$m$. The Rosseland optical depth reaches unity at $\log m = -0.9$. Here the
gravity amounts to \logg=5.8. Mass and electron densities are $6 \cdot
10^{-8}$~g\,cm$^{-3}$ and $5 \cdot 10^{15}$~cm$^{-3}$, respectively. Hence the
physical conditions in the line-forming regions are comparable to those in the
atmosphere of a subluminous B star. The vertical distribution of ionic fractions
of C and O is shown in Fig.\,\ref{fig_ion}. There we also show the H, He, and Ne
fractions that were taken from the test models that include these species in an
amount of 10\%. The dominant ionisation stages of oxygen at $\tau_{\rm Ross}=1$
are \ion{O}{ii} and \ion{O}{iii}, giving rise to prominent emission lines of
these ions in the spectra. In the case of carbon, \ion{C}{iii} dominates,
closely followed by \ion{C}{ii} and \ion{C}{iv}. This explains why we see lines
from three C ions. The dominant neon ions in the line-forming regions are
\ion{Ne}{ii} and \ion{Ne}{iii}.  For helium we have \ion{He}{ii} dominating,
followed by \ion{He}{i}, so we would expect strong lines from these ions if neon
and helium were abundant. Hydrogen is ionised by about 99.9\%, but
still, prominent emission lines are predicted if H were an abundant species (see
below).

\subsection{C and O lines compared to observations}

The synthetic spectrum is plotted together with the observed spectra in
Figs.~\ref{fig1} and \ref{fig2}. Let us first consider 4U\,0614+091. Generally,
many of the observed features are also seen in the model, although they do not match
the strength. This basically corroborates our line identifications. The
\ion{C}{ii} lines of the model are too weak (e.g.\ at 6580~\AA\ and 7235~\AA),
while the \ion{C}{iii} emissions are in some cases too weak (e.g.\ 4650~\AA) or
in other cases too strong (4180~\AA). The \ion{C}{iv}~5810~\AA\ line matches
well. The oxygen lines show a similar behaviour. The \ion{O}{ii} lines of the
model are too weak (e.g.\ at 4940~\AA).  Some lines of \ion{O}{iii} match
reasonably well (e.g.\ at 3720~\AA), while others do not (e.g.\ at 5590~\AA, which
is much too weak in the model).

The model comparison with 4U\,1626-67 confirms our ideas from the first
inspection of the spectrum. The lines in the blue region can be explained
qualitatively by the mere presence of oxygen lines. This, and the missing
\ion{C}{iv}~5810~\AA\ line might originate in a relatively low C/O ratio when
compared to 4U\,0614+091.

It is disappointing that our NLTE disk model obviously gives a poorer fit
to the observed line spectrum than the simple isothermal, constant-density, LTE
slab model presented by Nelemans \etal (2004). However, that slab model can at
best fit a a limited spectral range, as it is emitted from a particular emission
region with an assumed value for temperature and density and will never be able
to simultaneously fit the observed spectra from the X-ray to the UV and optical
ranges. In good agreement with the observation, our model exhibits lines from
three ionisation stages of carbon, indicating that temperature and density in
the line-forming regions are reproduced reasonably well. The fact that the
strength of many emission lines of a particular ion is either over- or
underestimated may stem from drawbacks in the model atoms. One reason could be
that the model ions are still too small and ignore the interlocking effects of
neglected energy levels. A further extension of the model ions is hampered by
the lack of atomic data, mostly oscillator strengths. Another reason for the
poor line fits could be errors in the electron collisional rates. Only few are
known from experiments or quantum mechanical calculations.

These problems also might affect our neon line-formation calculations. However,
the employed Ne model atom has been designed for the analysis of sdO stars and
gave successful fits to observed \ion{Ne}{ii} lines (Dreizler 1993). Therefore
we think that the predicted Ne emission lines in the model are more reliable
than the majority of the C and O lines.
Atomic data for H and He are accurately known and our model atoms have been
employed with success to analyse many classes of stellar spectra, hence,
we regard the computed line strengths of H and He as much more reliable than those
of the metal lines.
Another reason for the poor line fits might be that the assumed
  $\alpha$-disk does not describe the physics of the emitting region well.

\subsection{Limits on abundances of hydrogen, helium, and neon}

The upper abundance limits we derive for H and He are regarded as realistic
because of the above considerations, but the limit for neon is less secure.

In addition to the pure C/O disk model we also computed a model including 10\%
hydrogen and a model including 10\% helium in order to see if this allows the H
and He abundances to be constrained. In Fig.\,\ref{fig_hhe} we display the
resulting emission lines of H$_\beta$ and \ion{He}{ii}~5412~\AA. From the lack
of observed emission lines, it is evident that H and He are at most present at
the 10\% level. This excludes the possibility that the disks of 4U\,1626-67 and
4U\,0614+091 are dominated by hydrogen or helium and confirms the suspicion that
they are in fact C/O-dominated.

We also computed a C/O model that includes 10\% neon. It exhibits weak \ion{Ne}{ii}
lines,  but no such lines are detected in the VLT spectra
(Fig.\,\ref{fig_neon}).  We conclude that the neon abundance cannot be larger
than $\approx 10\%$.

\section{Summary}

We have presented new high-quality optical spectra of the ultracompact low-mass
X-ray binaries 4U\,0614+091 and 4U\,1626-67. They are pure emission line spectra
and most probably stem from the accretion disk. The spectral lines are
identified as due to \ion{C}{ii-iv} and \ion{O}{ii-iii}. Line identifications
are corroborated by first results from modeling the disk spectra with detailed
non-LTE radiation transfer calculations. Hydrogen and helium lines are lacking
and our models confirm the deficiency of H and He in the disk. Hence, the donor
stars in these systems are in fact the eroded cores of C/O white dwarfs. There
are indications that the O/C ratio in 4U\,0614+091 is higher than in
4U\,1626-67. This could suggest that the stripping process of the WD in
4U\,0614+091 is more advanced.

It is hard to estimate the systematic error of the derived upper limit for
the neon abundance so that the following conclusions are at the moment rather
uncertain. From the lack of \ion{Ne}{ii} lines we find that the Ne abundance
cannot exceed $\approx 10\%$ or, in other words, the Ne/O ratio is at most of
the order 0.2. The much higher Ne/O ratios ($\approx 0.7$) derived from the ISM
X-ray absorption edges of neutral Ne and O (Schulz \etal 2001, Juett \etal 2001)
would produce detectable \ion{Ne}{ii} lines in the disk spectra. This confirms
the conclusion of Juett \& Chakrabarty (2005), that the determined ISM
abundances of Ne and O are affected by ionisation effects and, hence, do not
reflect the abundances of the donor stars.

For an initial solar metallicity the $^{22}$Ne abundance in an Ne-enriched
crystallized and fractionated WD core can be estimated from theoretical models
to $\approx 0.07$, but this value is very uncertain (Yungelson \etal 2002). It
could be even higher by a factor of 3 (Isern \etal 1991).  Such an Ne-rich core
would have a mass of $\approx 0.06$~M$_\odot$ (Yungelson \etal 2002). The mass
of the WD donor in 4U\,1626-67 is much smaller (0.01~M$_\odot$; Yungelson \etal
2002). If we accept an upper limit of Ne=0.1 for the observed neon abundance,
then we may draw the following conclusion. Either the WD core has crystallized
and fractionated, then our observation favors a relatively small Ne-enrichment
as a result of the crystallization process. Or, if one accepts that
fractionation would result in a high Ne abundance of 0.2 (unobserved), then the
WD core in 4U\,1626-67 had no time to crystallize which, depending on various
details, lasts several Gyr (e.g.\ Hernanz \etal 1994).

Future work will concentrate on the disk modeling for the prototype 4U\,1626-67,
for which the observational database is the best of all such
systems. Ultimately, the spectral properties (flux distribution and emission
line strengths) over the complete wavelength range comprising spectral
observations with Chandra, HST, and VLT, must be explained by a unique disk
model.

\begin{acknowledgements}
T.R.\ is supported by the DLR under grant 50\,OR\,0201. We thank the referee for
constructive criticism that helped to improve the paper.
\end{acknowledgements}

\end{document}